\documentstyle[epsf]{mn}     

\newif\ifAMStwofonts

\ifoldfss 
  \newcommand{\rmn}[1] {{\rm #1}} 
  \newcommand{\itl}[1] {{\it #1}} 
  \newcommand{\bld}[1] {{\bf #1}} 
  \ifCUPmtlplainloaded \else 
    \NewTextAlphabet{textbfit} {cmbxti10} {} 
    \NewTextAlphabet{textbfss} {cmssbx10} {} 
    \NewMathAlphabet{mathbfit} {cmbxti10} {} 
    \NewMathAlphabet{mathbfss} {cmssbx10} {} 
  \fi 
  \ifAMStwofonts 
    \ifCUPmtlplainloaded \else
      \NewSymbolFont{upmath} {eurm10} 
      \NewSymbolFont{AMSa} {msam10} 
      \NewMathSymbol{\upi}     {0}{upmath}{19} 
      \NewMathSymbol{\umu}     {0}{upmath}{16} 
      \NewMathSymbol{\upartial}{0}{upmath}{40} 
      \NewMathSymbol{\leqslant}{3}{AMSa}{36} 
      \NewMathSymbol{\geqslant}{3}{AMSa}{3E} 
      \let\oldle=\le     \let\oldleq=\leq 
      \let\oldge=\ge     \let\oldgeq=\geq 
      \let\leq=\leqslant \let\le=\leqslant 
      \let\geq=\geqslant \let\ge=\geqslant 
    \fi 
  \fi 
\fi 
 
\ifnfssone 
  \newmathalphabet{\mathit} 
  \addtoversion{normal}{\mathit}{cmr}{m}{it} 
  \addtoversion{bold}{\mathit}{cmr}{bx}{it} 
  \newcommand{\rmn}[1] {\mathrm{#1}} 
  \newcommand{\itl}[1] {\mathit{#1}} 
  \newcommand{\bld}[1] {\mathbf{#1}} 
  \def\textbfit{\protect\txtbfit} 
  \def\textbfss{\protect\txtbfss} 
  \long\def\txtbfit#1{{\fontfamily{cmr}\fontseries{bx}\fontshape  
\long\def\txtbfit#1{{\fontfamily{cmr}\fontseries{bx}\fontshape{it}
    \selectfont #1}} 
  
\long\def\txtbfss#1{{\fontfamily{cmss}\fontseries{bx}\fontshape{n
}%
    \selectfont #1}} 
  \newmathalphabet{\mathbfit} 
\newmathalphabet{\mathbfit} 
  \addtoversion{normal}{\mathbfit}{cmr}{bx}{it} 
  \addtoversion{bold}{\mathbfit}{cmr}{bx}{it} 
  \newmathalphabet{\mathbfss} 
\textbfss{..} 
  \addtoversion{normal}{\mathbfss}{cmss}{bx}{n} 
  \addtoversion{bold}{\mathbfss}{cmss}{bx}{n} 
  \ifAMStwofonts 
    \ifCUPmtlplainloaded \else 
and 
your 
      \UseAMStwoboldmath 
      \makeatletter 
      \new@mathgroup\upmath@group 
      \define@mathgroup\mv@normal\upmath@group{eur}{m}{n} 
      \define@mathgroup\mv@bold\upmath@group{eur}{b}{n} 
      \edef\UPM{\hexnumber\upmath@group} 
      \new@mathgroup\amsa@group 
      \define@mathgroup\mv@normal\amsa@group{msa}{m}{n} 
      \define@mathgroup\mv@bold\amsa@group{msa}{m}{n} 
      \edef\AMSa{\hexnumber\amsa@group} 
      \makeatother 
      \mathchardef\upi="0\UPM19 
      \mathchardef\umu="0\UPM16 
      \mathchardef\upartial="0\UPM40 
      \mathchardef\leqslant="3\AMSa36 
      \mathchardef\geqslant="3\AMSa3E 
      \let\oldle=\le     \let\oldleq=\leq 
      \let\oldge=\ge     \let\oldgeq=\geq 
      \let\leq=\leqslant \let\le=\leqslant 
      \let\geq=\geqslant \let\ge=\geqslant 
    \fi 
  \fi 
\fi 
 
\ifnfsstwo 
  \newcommand{\rmn}[1] {\mathrm{#1}} 
  \newcommand{\itl}[1] {\mathit{#1}} 
  \newcommand{\bld}[1] {\mathbf{#1}} 
  \def\textbfit{\protect\txtbfit} 
  \def\textbfss{\protect\txtbfss} 
  
\long\def\txtbfit#1{{\fontfamily{cmr}\fontseries{bx}\fontshape{it}
    \selectfont #1}} 
  
\long\def\txtbfss#1{{\fontfamily{cmss}\fontseries{bx}\fontshape{n
}%
    \selectfont #1}} 
  \DeclareMathAlphabet{\mathbfit}{OT1}{cmr}{bx}{it} 
  \SetMathAlphabet\mathbfit{bold}{OT1}{cmr}{bx}{it} 
  \DeclareMathAlphabet{\mathbfss}{OT1}{cmss}{bx}{n} 
  \SetMathAlphabet\mathbfss{bold}{OT1}{cmss}{bx}{n} 
  \ifAMStwofonts 
    \ifCUPmtlplainloaded \else 
      \DeclareSymbolFont{UPM}{U}{eur}{m}{n} 
      \SetSymbolFont{UPM}{bold}{U}{eur}{b}{n} 
      \DeclareSymbolFont{AMSa}{U}{msa}{m}{n} 
      \DeclareMathSymbol{\upi}{0}{UPM}{"19} 
      \DeclareMathSymbol{\umu}{0}{UPM}{"16} 
      \DeclareMathSymbol{\upartial}{0}{UPM}{"40} 
      \DeclareMathSymbol{\leqslant}{3}{AMSa}{"36} 
      \DeclareMathSymbol{\geqslant}{3}{AMSa}{"3E} 
      \let\oldle=\le     \let\oldleq=\leq 
      \let\oldge=\ge     \let\oldgeq=\geq 
      \let\leq=\leqslant \let\le=\leqslant 
      \let\geq=\geqslant \let\ge=\geqslant 
    \fi 
  \fi 
\fi 
 
\ifCUPmtlplainloaded \else 
  \ifAMStwofonts \else 
    \def\upi{\pi} 
    \def\umu{\mu} 
    \def\upartial{\partial} 
  \fi 
\fi

   \title{ 
Calibration of the Faber-Jackson relation for M31 globular clusters
using Hipparcos data
}

   \author[Di Nella-Courtois H. et al.] 
{H. Di Nella-Courtois$^{1}$, P. Lanoix$^1$, G. Paturel$^1$\\
 $^1$CRAL-Observatoire de Lyon,
             F69561 Saint-Genis Laval, FRANCE\\
email : courtois@obs.univ-lyon1.fr} 
 
   \date{Received July 1998; accepted -- -- --}
\pubyear{1998}
 
\begin{document}
   \maketitle
 
   \begin{abstract}
    
In this paper we present a data analysis regarding globular clusters
as possible extragalactic distance indicators. For this purpose, we
collected all velocity dispersion measurements
published for galactic and M31 globular clusters.
The slope and the zero-point of the Faber-Jackson relation
were calibrated using  Hipparcos distance measurements, and 
the relation was applied to extragalactic globular clusters in M31. 
A distance modulus of 24.12 $\pm$ 0.45 mag was found. This is coherent 
with what is found by fitting the red giant branches of globular clusters
(24.47  $\pm$ 0.07, Holland 98),  and is found from
the peak of globular clusters luminosity function (24.03 $\pm$ 0.23, Ostriker and
Gnedin 97), but shorter than the 
24.7 $\pm$ 0.2 mag (Lanoix et al. 98) and 24.77 $\pm$ 0.11 mag
(Feast and Catchpole 97),  obtained by 
using Hipparcos data to calibrate the Cepheid period-luminosity.
This calibrated Faber-Jackson relation can now be directly use
for other Sc galaxies with resolved globular clusters, as soon as large amounts
of spectra will become available, e.g., through the VLT.

   \end{abstract}

\section{Introduction}
Since the 1970's much effort has been made to measure velocity
dispersions of globular clusters (gc's) in the Galaxy and later in M31 (Peterson 88).
Ten years after these pioneering measurements, new measurements have been published
using essentially 3m (Dubath and Grillmair 97) to 10m (Djorgovski et al. 97) telescopes.
Concordingly, the data of the Hipparcos satellite has been made available and
the distance of galactic gc's can be derived from parallaxes
independent of any standard candles.
Detailed correlations of gc's properties analogous to elliptical galaxies properties,
such as the Faber-Jackson (FJ) relation have been studied
for the Galactic system (Meylan and Mayor 86, Paturel and Garnier 92,
Fournier et al. 95, Djorgovski and Meylan 94, Djorgovski 95).
A recent  careful
analysis of these properties for the extragalactic system of M31 can
be found in Djorgovski et al. 97.
In particular we consider that  Djorgovski et al. 97 have demonstrated
that  M31 and the Galaxy gc's 
are similar systems in terms of metallicity.     
In agreement with all these studies, we show the validity of
a FJ for gc's, and propose a 
calibration of  the relation with the new Hipparcos data.\\

Section 2 describes the collection of data, section 3 includes the 
analysis and gives the calibration of the Faber-Jackson relation.
 In section 4, 
the calibrated relation is applied to M31 gc's and a
distance modulus is derived in agreement with recent independent measurements,
showing that this calibrated relation can now be applied to any
unbiased set of extragalactic gc's of an Sc host galaxy.

\section{The data}
\subsection{The galactic globular clusters}
\subsubsection{Data up to 1997}
In Table~\ref{gal.gcMW}, we present 
all the published measurements of the velocity dispersion
for 56 galactic gc's.
In Table~\ref{gal.gcMW}, one can find in columns  :
(1) NGC name,
(2) the integrated apparent V magnitude,
(3) the  absolute V magnitude,
(4) a raw average of all measurements of the velocity dispersion,
from the compilation of 38 references in Pryor and Meylan 93 (PRY93),
(5) the velocity dispersion from Dubath and Grillmair 97 
(DUB97), Zaggia 91 (ZAG91), Illingworth 76 (ILL76),
or an asterix if no measurement from an integrated light spectrum was available,
(6) the distance modulus.

Data in columns (2), (3), (6), were taken from the electronic version
dated 15th May 1997 of Harris 96 compilation. 

\subsubsection{Data from Hipparcos}
For 11 galactic gc's we found new distance measurements 
from the Hipparcos observations. 
They are shown in Table~\ref{hip.gcMW}.
In Table~\ref{hip.gcMW}, one can find in columns  :
(1) NGC name,
(2) Hipparcos distance modulus measurements with reference number, 
(3) average of Hipparcos distance modulus measurements,
(4) absolute V magnitude from Harris 97,
(5) calculated absolute V magnitude using Hipparcos data, see
section 3.2,
(6) an asterix if the available velocity dispersion comes from
individual star spectra.

\subsection{Andromeda globular clusters}

In Table~\ref{M31} we present the data concerning
29 gc's of M31
with a published measurement of their velocity dispersion.
Columns of Table~\ref{M31} correspond to: (1) Name from Sargent et al. 77,
(2) apparent V magnitude, (3) radial velocity,
(4) all velocity dispersion measurements with reference, (5) mean
velocity dispersion as used for the calculations in this paper.
Velocity dispersion measurements are taken from
Djorgovski et al. 97, Dubath and Grillmair 97, Dubath et al. 97, Peterson 88.
All apparent magnitudes and mean radial velocities come from Huchra
et al. 91.

\section{The analysis}
\subsection{The galactic globular clusters before Hipparcos}
Using the 56 galactic gc's with a measured velocity dispersion,
we performed a mean linear regression,
assuming the errors are both on the absolute magnitudes
and on the $log \sigma 's.$
We obtain with one cluster (NGC 2419) rejected at $3 \sigma$ :

\begin{equation}
M_{V}=(-4.00 \pm 0.33) log \sigma + (-4.71 \pm 0.27),
\end{equation}
the direct linear regression (assuming larger errors on dispersions
than on absolute magnitudes) gives:
\begin{equation}
M_{V}=(-3.29 \pm 0.33) log \sigma + (-5.24 \pm 0.27),
\end{equation}   
with r=0.81, r being the Pearson correlation factor, and
a dispersion around the FJ relation of Disp=0.68.

Considering that a globular cluster is constituted by some hundreds
of thousands of stars, we chose for a second analysis to eliminate 
the velocity dispersion
data originating from the measurements of singular star radial velocities.
As a matter of fact this kind of observations involve at the worst
around 10 stars and at the best around 150 stars. 
The selection of the stars for observation strategy involves choosing
bright stars or periferic ones. The selection criteria
for the observation of those stars will obviously affect the measurement
by adding biases such as the Malmquist one.\\
Eliminating gc's with a velocity dispersion measured from
individual stars spectra and keeping only the 
measurements obtained from integrated light spectra, we obtained a
subsample of 31 gc's.
We performed a mean linear regression,
and obtained with no cluster rejected at $3 \sigma$ :

\begin{equation}
M_{V}=(-4.12 \pm 0.57) log \sigma + (-4.49 \pm 0.51),
\end{equation}
the direct linear regression gives :
\begin{equation}
M_{V}=(-3.08 \pm 0.57) log \sigma + (-5.39 \pm 0.51),
\end{equation} 
with r=0.71, r being the Pearson correlation factor, and
a dispersion around the FJ relation of Disp=0.73.

\begin{figure}
\epsfxsize=8cm
\hbox{\epsfbox[50 410 510 740]{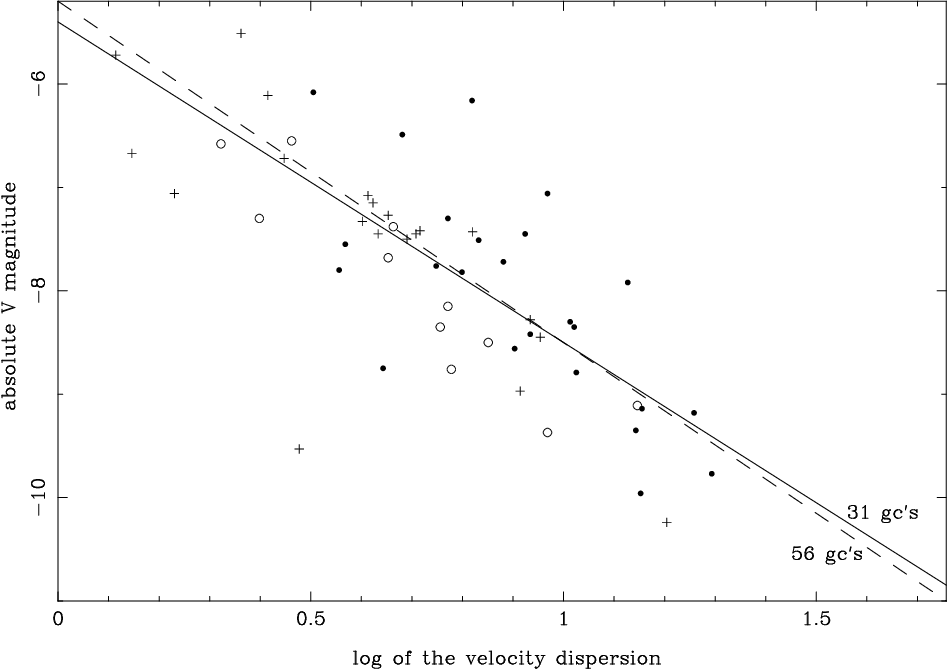}}
\caption{
The Faber-Jackson relation for 56 galactic globular clusters,
obtained with all the measurements (dashed line) and with only
velocity dispersions (31 gc's) measured from an integrated light spectrum
(solid line).
The 11 gc's 
which will be calibrated by Hipparcos later in the paper are
in open circles, the remaining 21 gc's with individual star measurement
are plotted as crosses and the  remaining 24  gc's with a velocity 
dispersion measured  from an integrated light spectrum
are plotted in filled circles.
}
\label{FJ1}
\end{figure}

One can see on Figure~\ref{FJ1} the mean FJ relations for the 56 gc's
obtained with all  measurements (dashed line) and with only
velocity dispersions (31 gc's) measured from an integrated light spectrum
(solid line).

Excluding the 7 gc's calibrated by Hipparcos, the direct regression
on the 24 remaining gc's gives, with no rejection :

\begin{equation}
M_{V}=(-3.33 \pm 0.71) log \sigma  + (-5.03 \pm 0.66),
\end{equation}   
 with r=0.71, and  Disp=0.75.

\subsection{The FJ calibration from globular clusters measured by Hipparcos}

For 11 gc's with a new distance determination obtained
by Hipparcos, we can re-calculate their absolute magnitude. 
In order to take into account the same extinction correction as in Harris 97 on the
apparent magnitudes, we recalculated the absolute magnitude in V  using:

\begin{equation}
M_{V}post = -dist. mod. Hip. + 5 (log d) - 5 + M_{V}pre,\\
\end{equation}   
 with $d$ and $M_{V}pre$ from Harris 97 as in  Table~\ref{gal.gcMW},
and dist. mod. Hip. as in the third column of  Table~\ref{hip.gcMW}.
The updated $M_{V}post$ magnitude is found in the fifth column 
of Table~\ref{hip.gcMW}.\\

According to Hipparcos the clusters are systematically further away 
than thought from previous measurements. 
{\it In the mean we observe a shift of 0.34 mag between
pre and post-Hipparcos measurements.}
We performed a direct linear regression on the 11 gc's, assuming larger errors
on the $ log\sigma$'s than on the absolute magnitude,
and obtain with no cluster rejected at $3 \sigma$ :

\begin{equation}
M_{V}pre = (-3.59 \pm 0.50) log \sigma  + (-5.44 \pm 0.38),\\
\end{equation}
with r=0.92  and Disp=0.39.
\begin{equation}
M_{V}post = (-3.46 \pm 0.47) log \sigma  + (-5.88 \pm 0.35),\\
\end{equation}
with r=0.93  and Disp=0.37.

7 out of these 11 clusters have a velocity dispersion measured from an
integrated light spectrum, we obtain for these 7 clusters,
with zero rejection at  $3 \sigma$ , the direct regression :

\begin{equation}
M_{V}pre = (-3.58 \pm 0.66) log \sigma + (-5.48 \pm 0.52),
\end{equation}
with r=0.92  and Disp=0.42.   
\begin{equation}
M_{V}post = (-3.39 \pm 0.58) log \sigma + (-6.01 \pm 0.46),
\end{equation}
with r=0.93 and Disp=0.37.

\begin{figure}
\epsfxsize=8cm
\hbox{\epsfbox[50 160 500 480]{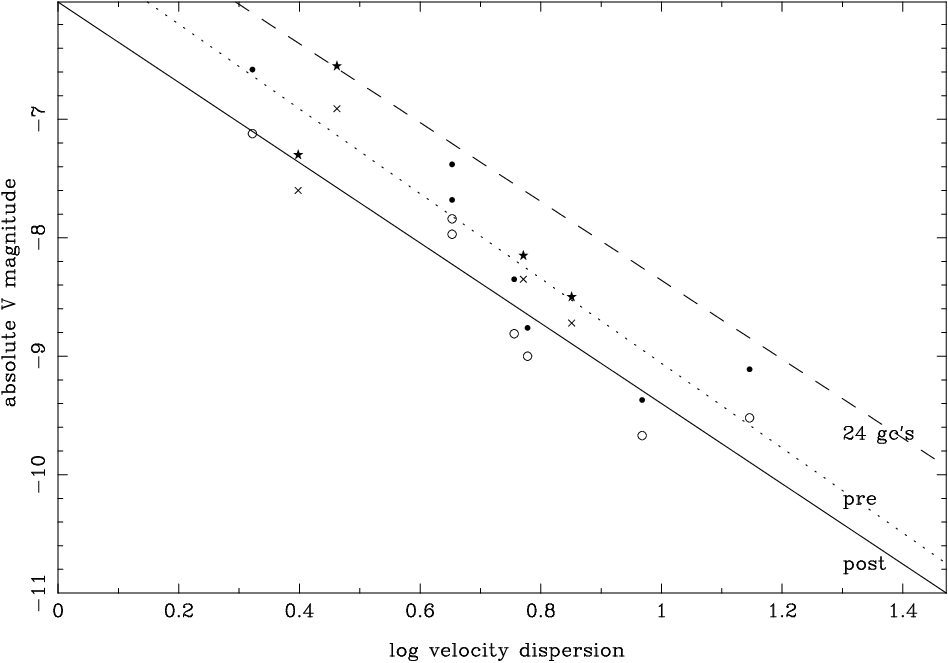}}
\caption{
For the 7 gc's calibrated by Hipparcos and with a measurement of the
velocity dispersion from an integrated light spectrum,
one can see the pre-Hipparcos
absolute magnitudes (filled circles) and the post-Hipparcos ones (open circles).
For the 4 remaining gc's  calibrated by Hipparcos  and with a measurement of the 
velocity dispersion from singular star spectra
one can see the pre-Hipparcos
absolute magnitudes (filled stars)  and the post-Hipparcos ones (crosses).
The solid line represents the direct post-Hipparcos Faber-Jackson relation for 7 gc's both
measured by Hipparcos and with velocity dispersions measured 
from an integrated light spectrum, the dotted line is the direct pre-Hipparcos
relation.
The dashed line reminds us 
of the direct FJ relation found for the subsample
of 24 gc's studied previously.
We observe a systematic shift towards brighter magnitudes  on the zero points
while the slopes are not significantly modified.
}
\label{FJ2}
\end{figure}

One can see in Figure~\ref{FJ2} the direct pre-Hipparcos FJ relation  
obtained with only
velocity dispersions measured from an integrated light spectrum (equation 9)
in solid line and the direct post-Hipparcos (equation 10) in doted line.
The dashed line reminds us of the direct FJ relation found for the subsample
of 24 gc's studied previously (equation 5) excluding these former 7 gc's.

From Figure~\ref{FJ2} a clear shift (about 1 mag) in  the
zero point between the 24 gc's sample and the
post-Hipparcos one is noted,  while the slope is not significantly modified.
But only part of the offset arises from the Hipparcos result.

We have already seen in Figure~\ref{FJ1} that the gc's measured by Hipparcos
lie systematically below the fits of the FJ relations and in  Figure~\ref{FJ2},
this is explicitely shown. We calculated an average 
difference of 0.55 mag between the FJ relation
fitted on the 24 gc's and the pre-Hipparcos absolute magnitudes of the
7 gc's. Obviously the Hipparcos observations were dedicated to intrinsic
bright gc's.

As we noted previously, the difference between the pre and post-Hipparcos
absolute magnitudes is in the mean of 0.34 mag.

This means the real offset coming from the Hipparcos measurements is not of
1 mag but a shift of 0.34 mag on the 24 gc's.
After shifting these gc's towards brighter absolute magnitudes, we
obtain the calibrated direct FJ relation, (31 calibrated gc's):

\begin{equation}
M_{V}FJ = (-3.0  \pm 0.3) log \sigma + (-5.8 \pm 0.1)
\label{FJ}
\end{equation}

\subsection{Andromeda globular clusters}

From Table~\ref{M31} we eliminated 3 gc's 
: M31-279, M31-315, M31-090 for which no
reliable measurement of the velocity dispersion is available. 

Using the remaining 26 values of velocity dispersions, we looked
for a correlation with the {\it apparent V} magnitude, taking
into account that all these clusters are approximately at the same
distance from the observer. If one finds a slope in apparent
magnitude in agreement with the slope in absolute magnitude obtained
for gc's in the Galaxy, one could suppose the sample 
isn't affected by a Malmquist bias regarding the selection of the
extragalactic gc's.
We obtained the best mean regression fit, after rejecting M31-144
and M31-219 at 3$\sigma$:

\begin{equation}
m_{V}=(-4.2 \pm 0.4) log \sigma + (20.0 \pm 0.5)\\
\end{equation}                                      
the direct linear regression gives :
\begin{equation}
M_{V}=(-3.7 \pm 0.4) log \sigma  + (19.4 \pm 0.5),\\
\end{equation}                                      
with r=0.87, Disp=0.34.\\

The slopes are quite coherent with the slopes obtained for the
galactic globular clusters (equations 1-5) and the differences
of slopes are included in the error bars.
The direct FJ relation for M31 gc's is also coherent with the 
direct FJ relation calibrated from Hipparcos (equations 7-10).
We conclude that this sample of extragalactic gc's 
can be considered as free from bias (although it is not complete
of course).
 And thus we can apply our calibrated galactic 
direct Faber-Jackson relation to this sample.

We should also note that the velocity dispersions 
measured in  M31 gc's
are systematically larger than for the ones given in our
galaxy. This is due to an observational selection effect, the 
intrinsic bright gc's which are easier to observe
from a distance, have a larger velocity dispersion.
From various comparative studies of the galactic and the M31
gc's systems, in particular on the metallicity, we can suppose
the two systems are globally comparable.
With the Very Large Telescope, one will be able to measure a
larger sample of gc's in M31 and to compare the distribution
in velocity dispersions.\\

\begin{figure}
\epsfxsize=8cm
\hbox{\epsfbox[50 160 500 480]{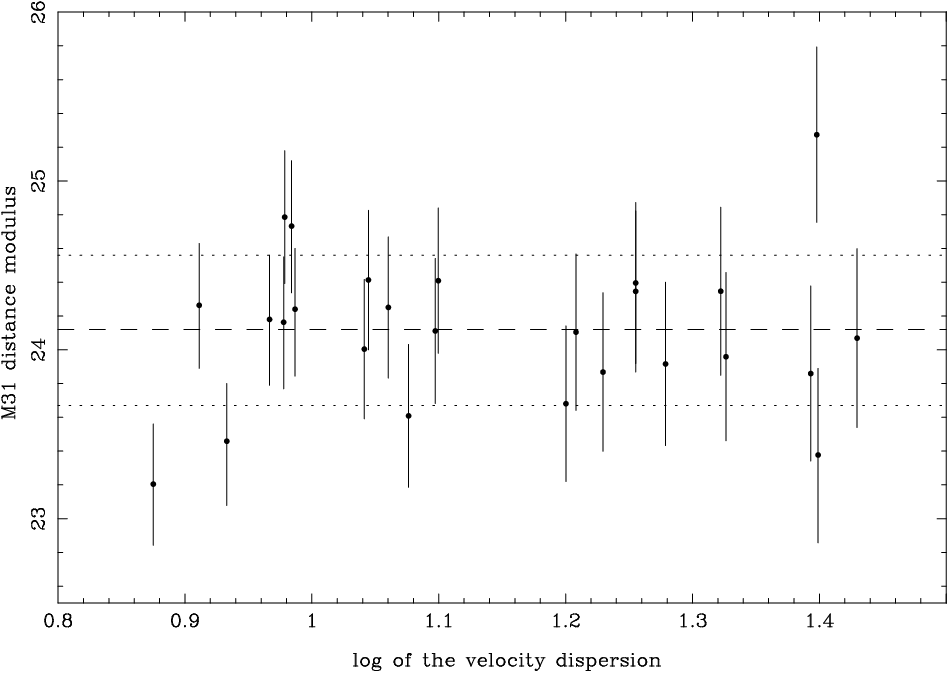}}
\caption{
M31 distance modulus versus $log \sigma $, obtained from
equation~\ref{FJ} and corrected for absorption,  
for 26 globular clusters. 
}
\label{FJ3}
\end{figure}

We considered a foreground reddening of 0.1 mag (Frogel
et al. 80). The absorption was taken to be 3.2 times the
reddening (Da Costa and Armandroff 90).\\
Applying equation~\ref{FJ} to our sample we obtain the
distance moduli shown in  Figure~\ref{FJ3}.
The resulting mean distance modulus for M31 is : 24.12.
If we use the extremes given by the error bars on the slope
and zero point of equation~\ref{FJ} we obtain a mean error on the distance 
modulus of $\pm$ 0.45 mag
In  Figure~\ref{FJ3} we draw the error bars for each gc given by
the lowest slope and zero point and by the highest ones.
The huge error bar on the determination of the distance modulus
is directly due to the small numbers of objects involved in this analysis and
large dispersions on the FJ relations.

One can see in  Figure~\ref{FJ3} that despite the poor common range of velocity
dispersions between the galactic and M31 gc's available (0.85 to 1.15), there is no
systematic effect seen towards larger dispersions on the calculated
distance moduli.
This implies that the slope of the FJ relation is similar to the one
in our galaxy, as we suggested previously from the study of the apparent
magnitudes versus the velocity dispersions.

\section{Conclusion}
In this paper, we present an analysis of globular cluster velocity dispersions
as possible distance indicators. Using Hipparcos recent set of distance measurements
published for 11 galactic gc's, we give a calibration of the
Faber-Jackson relation for gc's. This calibration is used on 26
gc's in M31, to derive a 
mean distance modulus  24.12 $\pm$ 0.45 mag.
The value we find is coherent with 
what is found by fitting the red giant branches of gc's 
(24.47 $\pm$ 0.07,  Holland 98), and found from
the peak of gc's luminosity function ( 24.03 $\pm$ 0.23, Ostriker and
Gnedin 97), but shorter than  the 
24.7 $\pm$ 0.2 mag (Lanoix et al. 98) and
24.77 $\pm$ 0.11 mag (Feast and Catchpole 97), obtained by
using Hipparcos data to calibrate the Cepheid period-luminosity.
We also demonstrated that this calibration can be used for extragalactic 
gc's systems in an Sc host galaxy even though it's accuracy of 0.45 mag is not so good.
The huge error bar on the determination of the distance modulus
is directly due to the small numbers of objects involved in this analysis and
large dispersions on the FJ relations. This shows a need to enlarge the
set of galactic and M31 gc's measured in velocity dispersions.
                                          
Future measurements with the VLT for galaxies as far as Virgo
will be ready for use with this calibration independent of any standard candles.

\begin{table*}
\caption{
Galactic globular clusters } 
\label{gal.gcMW}
\begin{tabular}{llllll}
\hline
NGC & $m_{Vt}$ &  $M_Vt$ & $\sigma$ km/s & $\sigma$ km/s from        & $R_{sun} $    \\
    &          &         & raw average    & integrated light spectra    & kpc  \\
    &          &         & from PRY93     &  or * = from star spectra   &            \\
(1) & (2)      & (3)     & (4)            &  (5)                        & (6)        \\
\hline
 104 & 3.95 & -9.37& 11.5 & 9.3 DUB97&4.3  \\ 
 288 & 8.09 & -6.55& 2.9  & *     & 8.1  \\      
 362 & 6.40 & -8.35& 6.4  & 5.7 DUB97&  8.3\\
1851 & 7.14 & -8.35& 10.4 & 10.5 DUB97& 12.2\\
1904 & 7.73 & -7.80& 5.2  & 3.6 DUB97&12.6\\
2419 &10.39 & -9.53& 3.0  & *     &82.3\\
2808 & 6.20 & -9.35& 13.4 & 13.9 ZAG91 & 9.3\\
3201 & 6.75 & -7.42&  5.2 & *     & 5.1\\
4147 &10.32 & -6.11&  2.6 & *     &18.8 \\
4590 & 7.84 & -7.30&  2.5 & *     &10.1 \\
5053 & 9.47 & -6.67& 1.4  & *     &16.2\\
5139 & 3.68 &-10.24& 16.0 & *     & 5.1\\
5272 & 6.30 & -8.75& 5.6  & 4.4 DUB97&10.0 \\
5286 & 7.34 & -8.56& 8.0  & 8.0 DUB97&10.7 \\
5466 & 9.04 & -7.06& 1.7  & *     & 16.6\\
5694 &10.17 & -7.76& 5.5  & 5.6 DUB97&33.9\\
5824 & 9.09 & -8.79& 11.6 & 10.6 DUB97&31.3\\
5904 & 5.65 & -8.76& 5.7  & 6.0 DUB97& 7.3\\
5946 & 9.61 & -7.55& 3.7  & 3.7 DUB97&12.3\\
6093 & 7.33 & -7.92& 12.4 & 13.4 DUB97& 8.7\\
6121 & 5.63 & -7.15& 4.2  & *     &  2.2\\
6171 & 7.93 & -7.08& 4.1  & *     &  6.3\\
6205 & 5.78 & -8.50& 7.1  & *     &  7.0\\
6218 & 6.70 & -7.27& 4.5  & *     &  4.7\\
6254 & 6.60 & -7.43& 6.6  & *     &  4.3\\
6256 &11.29 & -6.16& 6.5  & 6.6 DUB97&  9.3 \\
6266 & 6.45 & -9.14& 14.3 & 14.3 DUB97&  6.7 \\
6284 & 8.83 & -7.82& 6.2  & 6.3 DUB97& 14.3 \\
6293 & 8.22 & -7.72& 7.6  & 7.6 DUB97&  8.8 \\
6325 &10.33 & -7.30& 5.8  & 5.9 DUB97&  9.4 \\
6341 & 6.44 & -8.15& 5.9  & *     &  8.1 \\
6342 & 9.66 & -6.49& 4.6  & 4.8 DUB97&   9.1 \\
6366 & 9.20 & -5.72& 1.3  & *     &   3.6\\
6362 & 7.73 & -6.72& 2.8  & *     &   7.5\\
6388 & 6.72 & -9.77& 18.9 & 18.9 ILL76&  11.5\\
6397 & 5.73 & -6.58& 4.5  & 2.1 DUB97&   2.2 \\
6441 & 7.15 & -9.18& 18.0 & 18.1 DUB97&   9.7 \\
6522 & 8.27 & -7.51& 6.7  & 6.8  DUB97&   7.0 \\
6535 &10.47 & -4.68& 2.4  & *     &   6.8\\
6541 & 6.30 & -8.42& 8.2  & 8.6 ZAG91&  7.4\\
6558 & 9.26 & -6.08& 2.9  & 3.2 DUB97&  6.4 \\
6624 & 7.87 & -7.45& 5.4  & 8.4 ZAG91 &  7.9\\
6626 & 6.79 & -8.28& 8.6  & *     &   5.7\\
6656 & 5.10 & -8.45& 9.0  & *     &   3.2\\
6681 & 7.87 & -7.06& 5.1  & 9.3 DUB97&   8.7 \\
6712 & 8.10 & -7.45& 4.3  & *     &   6.7\\
6715 & 7.60 & -9.96& 14.2 &  14.2 ILL76 & 26.2\\
6752 & 5.40 & -7.68& 4.5  & 4.5 DUB97&  3.9 \\
6779 & 8.37 & -7.33& 4.0  & *     &   9.9\\
6809 & 6.32 & -7.50& 4.9  & *     &   5.3\\
6838 & 8.19 & -5.51& 2.3  & *     &   3.8\\
6864 & 8.52 & -8.30& 10.3 & 10.3 ILL76& 18.4\\
6934 & 8.83 & -7.45& 5.1  & *     &  15.2\\
7078 & 6.20 & -9.11& 12.0 & 14.0 DUB97& 10.2 \\
7089 & 6.47 & -8.97& 8.2  & *     &  11.4\\
7099 & 7.19 & -7.38& 5.6  & 4.6 DUB97&  7.9 \\
\hline
\end{tabular}
\end{table*}

\begin{table*}
\caption{
New distance determinations from Hipparcos. References: (1) Gratton et al. 97a, (2)
Bartkevicius et al. 97, (3) Reid in Heber et al. 97, (4) Gratton et al. 
in Heber et al. 97, (5) Pont et al. 97.} 
\label{hip.gcMW}
\begin{tabular}{llllll}
\hline
NGC &  Hipparcos  distance moduli & average & pre-Hipparcos $M_{V}$ & post-Hipparcos $M_{V}$ & * = star spectra\\
 (1)&     (2)                     & (3)     & (4)               &  (5)       & (6)          \\
\hline
 104 & 13.63 (1) 13.3 (2) & 13.47  &-9.37 &-9.67 & \\
 288 & 14.95 (1) 15.00 (3) 14.76 (4)& 14.90 &-6.55 &-6.91 &* \\    
 362 & 15.06 (1) & 15.06& -8.35&-8.81& \\
4590 & 15.32 (1) & 15.32&-7.30&-7.60& *\\
5904 & 14.61 (1) 14.53 (3) 14.58 (4) 14.5 (2) & 14.56&-8.76&-9.00& \\
6205 & 14.45 (1) & 14.45&-8.50&-8.72& *\\
6341 & 14.61+-0.08 (5) 14.81 (1) 14.93 (3) 14.83 (4) & 14.74&-8.15 & -8.35 & *\\
6397 & 12.25 (3) & 12.25&-6.58& -7.12 &\\
6752 & 13.32 (1) 13.17 (3) 13.20 (4) & 13.25 & -7.68& -7.97&\\
7078 & 15.45 (3) & 15.45&-9.11 &-9.52 &\\
7099 & 14.95 (1) & 14.95 &-7.38 & -7.84&\\
\hline
\end{tabular}
\end{table*}

\begin{table*}
\caption{
Andromeda globular clusters} 
\label{M31}
\begin{tabular}{lllll}
\hline
Sargent et al.  77 & $m_{V}$    & mean Vr km/s  & various $\sigma$ km/s  &mean $\sigma$ km/s    \\
   (1)     &  (2) &   (3)         &        (4)            &  (5)                \\                  &      \\
\hline
M31-001&  13.70   &-331.0   &25.06 Dj &   25.06    \\
M31-002&  15.80   &-380.0   &9.70 Dj  &    9.70    \\
M31-058&  15.80   &-226.3   &10.60 Du 11.56 Dj &   11.08   \\  
M31-064&  15.00   &-373.0   &16.15 Dj &   16.15    \\
M31-072&  14.60   &-210.5   &19.00 Pe &  19.00   \\
M31-073&  14.60   &-350.8   &18.00 Pe 14.27 Dj 15.3 Du &   15.86  \\
M31-078&  14.20   &-414.0   &24.00 Pe 25.46 Dj &   24.73  \\
M31-090&  16.70   &-412.0   &$<$10.00 &   10.00  \\
M31-105&  16.30   &-400.8   &9.08 Dj 10.20 Du &    9.64   \\
M31-108&  15.80   &-404.0   &9.82 Dj 8.70 Du &    9.26   \\
M31-144&  15.60   &-344.0   &25.00 Pe &  25.00   \\
M31-199&  15.40   & -88.0   &11.00 Pe *6.00 Du &  11.00  \\
M31-213&  14.50   &-186.5   &40.00 Pe 21.90 Du 20.50 Dj &   21.20  \\
M31-217&  14.90   &-161.0   &21.00 Pe &  21.00 \\
M31-219&  15.10   &-292.0   &7.10 Du 8.11 Dj &    7.50  \\
M31-222&  15.10   &-241.0   &18.00 Pe &  18.00  \\
M31-233&  15.15   &-325.0   &18.00 Pe &  18.00  \\
M31-244&  15.34   & -53.0   &12.00 Pe 13.20 Dj &   12.51 \\
M31-272&  14.70   &-215.2   &16.30 Du 17.62 Dj &   16.96 \\
M31-279&  15.35   &-131.0   &$<$10.00 Pe & 10.00 \\
M31-280&  14.30   &-164.8   &26.90 Du $>$40 Pe &   26.90  \\
M31-302&  14.90   &  -8.0   &11.92 Dj &   11.92     \\
M31-305&  15.63   &-205.0   &12.58 Dj &   12.58   \\
M31-312&  16.05   &-352.0   &8.15 Dj&    8.15    \\
M31-315&  15.60   & -95.0   &$<$10.00 Pe & 10.00  \\
M31-319&  15.75   &-360.0   &9.10 Du 10.10 Dj &    9.50  \\
M31-322&  15.59   &-349.0   &11.49 Dj &   11.49  \\
M31-351&  15.18   &-208.0   &8.57 Dj &    8.57    \\
M31-352&  16.37   &-326.0   &9.52 Dj &    9.52    \\
\hline
\end{tabular}
\end{table*}

\section*{Acknowledgments}We are grateful to R. Garnier, P. Dubath
and the referee for their help.

\end{document}